\pdfoutput=1

\documentclass{article}
\usepackage[utf8]{inputenc}
\usepackage[a4paper, margin=25mm]{geometry}

\usepackage[fontsize=10.5pt]{fontsize}
\usepackage{mathptmx}
\usepackage{helvet}

\usepackage{amsfonts,amsmath,amsthm,amssymb}
\usepackage{bm}
\usepackage{comment}
\usepackage[sort&compress, numbers, round]{natbib}

\usepackage{lipsum}

\usepackage[colorlinks=true, urlcolor=blue]{hyperref}
\usepackage{cleveref}
\usepackage{afterpage}
\usepackage{graphicx}

\usepackage{xcolor}
\definecolor{headertextcolor}{RGB}{89,89,89}
\definecolor{perclr}{RGB}{77, 175, 74}
\definecolor{medclr}{RGB}{152, 78, 163}

\usepackage{authblk}

\usepackage{manyfoot}
\DeclareNewFootnote{B}[arabic]

\usepackage{indentfirst}

\usepackage[explicit, largestsep]{titlesec}

\titleformat{\section}[block]
    {\fontsize{12}{14.4}\selectfont\sffamily\bfseries}
    {\makebox[1cm][l]{\thesection.}\MakeUppercase{#1}}{0pt}{}
\titleformat{name=\section, numberless}[block]
    {\fontsize{12}{14.4}\selectfont\sffamily\bfseries}
    {\MakeUppercase{#1}}{0pt}{}
\titlespacing*{\section}{0pt}{12pt}{1pt}

\titleformat{\subsection}[block]
  {\selectfont\sffamily\bfseries}
  {\makebox[1cm][l]{\thesubsection}#1}{0pt}{}
\titlespacing*{\subsection}{0pt}{6pt}{\parskip}
 
\titleformat{\subsubsection}[block]
  {\selectfont\sffamily\bfseries}
  {\makebox[1cm][l]{\thesubsubsection}#1}{0pt}{}
\titlespacing*{\subsubsection}{0pt}{3pt}{\parskip}

\usepackage{fancyhdr}
\fancypagestyle{firststyle}{
    \fancyhf{}
    \fancyhead[C]{%
        \begin{minipage}{1.85in}%
            \includegraphics[width=1.63in]{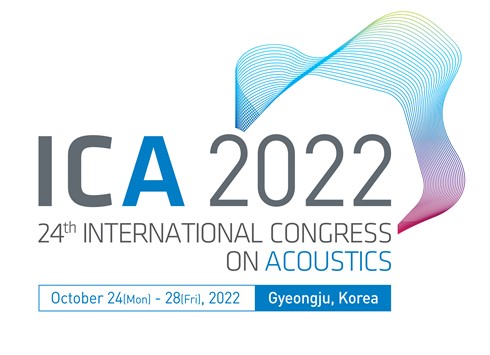}\\
        \end{minipage}
        \begin{minipage}[c]{3.34in}%
            \fontsize{10}{12}\selectfont\sffamily\bfseries
            \color{headertextcolor}
            PROCEEDINGS of the\\
            24th International Congress on Acoustics\\\\
            October 24 to 28, 2022 in Gyeongju, Korea
        \end{minipage}
    }

    \fancyfoot[C]{%
        \includegraphics[width=1.1in]{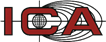}
        \hskip0.5in
        \includegraphics[width=0.53in]{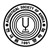}
    }

}
\makeatletter
\renewcommand{\maketitle}{
    \vspace*{.75in}
    \begin{center}
        \fontsize{14}{16.8}\selectfont\sffamily\bfseries
        \@title
    \end{center}
    \begin{center}
        \@author
    \end{center}
    \enlargethispage{-0.25in}
}
\makeatother

\makeatletter
\renewcommand{\abstractname}{{\sffamily\bfseries ABSTRACT\\}}
\renewcommand{\abstract}{\noindent\abstractname}
\g@addto@macro\endabstract{\vspace{1em}\par}
\makeatother

\newcommand{\keywords}[1]{\noindent Keywords: #1}

\makeatletter
\def\@makecaption#1#2{%
  \vskip\abovecaptionskip
  \hb@xt@\hsize{\hfil#1 -- #2\hfil}%
  \vskip\belowcaptionskip}
\makeatother

\usepackage{tabularx}
\usepackage{multirow}
\usepackage{booktabs}

\usepackage{siunitx}

\usepackage[style=base]{caption}

\begin{document}

\title{Do uHear? Validation of uHear App for Preliminary Screening of Hearing Ability in Soundscape Studies}

\author[1,*]{Zhen-Ting ONG}
\affil[1]{School of Electrical and Electronic Engineering, Nanyang Technological University, Singapore}
\author[1,*]{Bhan LAM}
\author[1]{Kenneth OOI}
\author[1]{\\Karn N. WATCHARASUPAT}
\author[1]{Trevor WONG}
\author[1]{Woon-Seng GAN}
\FootnotetextB{1}{\{ztong, bhanlam, wooi002, karn001, trevor.wong, ewsgan\} @ ntu.edu.sg}
\FootnotetextB{*}{These authors contributed equally to this work}

%\affil[2]{Institution, Country}
%\FootnotetextB{2}{gf@bbb.co.jp}

%\affil[3]{Institution, Country}
%\FootnotetextB{3}{gf@ccc.go.jp [Please note: It is optional to provide the email address(es) of the author(s). Please make sure that your co-authors concur with the mention of their email address in this paper.]}

\maketitle

\begin{abstract}
    Studies involving soundscape perception often exclude participants with hearing loss to prevent impaired perception from affecting experimental results. Participants are typically screened with pure tone audiometry, the ``gold standard'' for identifying and quantifying hearing loss at specific frequencies, and excluded if a study-dependent threshold is not met. However, procuring professional audiometric equipment for soundscape studies may be cost-ineffective, and manually performing audiometric tests is labour-intensive. Moreover, testing requirements for soundscape studies may not require sensitivities and specificities as high as that in a medical diagnosis setting. Hence, in this study, we investigate the effectiveness of the uHear app, an iOS application, as an affordable and automatic alternative to a conventional audiometer in screening participants for hearing loss for the purpose of soundscape studies or listening tests in general. Based on audiometric comparisons with the audiometer of 163 participants, the uHear app was found to have high precision (\SI{98.04}{\%}) when using the World Health Organization (WHO) grading scheme for assessing normal hearing. Precision is further improved (\SI{98.69}{\%}) when all frequencies assessed with the uHear app is considered in the grading, which lends further support to this cost-effective, automated alternative to screen for normal hearing.
    % Abstract must be 100-200 words.
\end{abstract}

\afterpage{\aftergroup\restoregeometry}
% NOTE: MAX 8 PAGES!!!

\section{Introduction}

In the normative minimum reporting requirements of soundscape studies, the hearing ability of all human participants must be explicitly reported \citep{InternationalOrganizationforStandardization2018}. Moreover, the required computation of psychoacoustic loudness of the acoustic environment represents an ``average loudness as perceived by a group of individuals with otologically normal hearing'' \citep{InternationalOrganizationforStandardization2017a}. Despite the inherent requirements to determine normal hearing ability in listening experiments, there are no explicit guidelines to grade the ``normality'' of hearing ability even in technical directives \citep{InternationalOrganizationforStandardization2018,InternationalTelecommunicationUnionRadiocommunicationSector2015}. Moreover, it may also be of interest to investigate the soundscape perception of persons with varying degrees of age-related hearing loss (i.e. elderly). Therefore, there is a significant impetus for an automated audiometric test that does not require expert knowledge or professional audiometric equipment for the purposes of screening ability before listening experiments. 

Hearing ability is usually assessed by audiometric tests administered by a trained healthcare professional (e.g. audiologists, otolaryngologists). Since hearing impairment is a medical diagnosis, audiometric tests are manually administered by qualified personnel and governed by internationally recognised guidelines \citep{ASHA2005,BSA2018,ISOAudiometry2010}, and conducted in strict acoustic conditions with calibrated audiometric equipment \citep{InternationalOrganizationforStandardization2018a}. Beyond medical diagnoses, compliance with the stringent requirements of audiometric tests to assess hearing ability for listening experiments becomes prohibitively expensive and laborious. Nonetheless, a number of soundscape studies with listening tests have reported the hearing ability of the recruited participants based on pure tone audiometry (PTA) with professional audiometric equipment \citep{Hong2020b, Medvedev2015, Hao2016, Lee2015,RadstenEkman2015a,Axelsson2010}.

To overcome the barriers due to cost and availability of audiometric tests and qualified healthcare personnel, mobile health technologies have been developed and deployed, especially in the form of smartphone-based apps for preventive screening in schools and for the elderly \citep{WHOscreening2021,Manus2021}. A recent meta-analysis of smartphone-based audiometry reported a respectable aggregated diagnostic sensitivity of \SI{89}{\%} [(\SI{83}{\%}, \SI{93}{\%}), \SI{95}{\%} CI] and specificity of \SI{93}{\%} [(\SI{87}{\%}, \SI{97}{\%}); \SI{95}{\%} CI] for PTA \citep{Chen2021}. Few apps have been scientifically validated, with the rarity of the iOS-based uHear app (uHear, Unitron -- Sonova Holding AG, Stäfa, Switzerland) subjected to some rigor \citep{Saliba2017,Li2020b,Melo2022,Barczik2018}. Moreover, it was found that device output levels were similar across mobile phone models running the iOS platform (Apple Inc., California, USA) \citep{Patel2021}, and its bundled headphones were found to be more accurate than supra- or circumaural headphones \citep{Barczik2018}, thereby increasing the availability of any iOS-based audiometry apps. To the authors' best knowledge, there are also no published soundscape literature that screened for hearing ability using smartphone-based applications.

To this end, there is an absence of studies examining the efficacy of smartphone-based audiometric apps to primarily screen for normal hearing. Based on reported literature, this study thus examines the efficacy of the iOS-based uHear app in comparison with a ``gold standard'' audiometer in detecting normal hearing based on the latest World Health Organisation (WHO) grading scheme \citep{WorldHealthOrganisation2021}. 

\section{Methodology}

\subsection{Participants}

This study was conducted as an auxiliary to two ongoing subjective listening experiments at Nanyang Technological University (NTU), Singapore. A total of 163 participants were recruited after first screening for any self-reported hearing loss, of which there were none. Participants were approximately evenly distributed in gender [female: 91 (\SI{55.8}{\%}); male: 72 (\SI{44.2}{\%})] with a relatively young age distribution between 18 and 75 years ($\Bar{x}_\text{age}=27.3$ years, $\text{SD}_\text{age}=11.4$, $\text{SE}_\text{age}=0.89$). Formal ethical approvals were obtained from the Institutional Review Board (IRB) of NTU (Ref. IRB-2020-08-035 and IRB-2021-293) for the responses collected. 

%We recruited 163 participants via study from Affective Responses to Augmented Urban Soundscapes (ARAUS) and Soundscape Attributes Translation Project (SATP). The age distribution for mean was 27.29 and standard distribution (SD) was 11.39. The gender distribution was 91 female and 72 male. No self-reported hearing loss for all participants. 
% (since it's part of ARAUS study anyway).

\subsection{Pure Tone Audiometry}

\begin{figure}[t]
    \centering
    \includegraphics[width=0.9\textwidth]{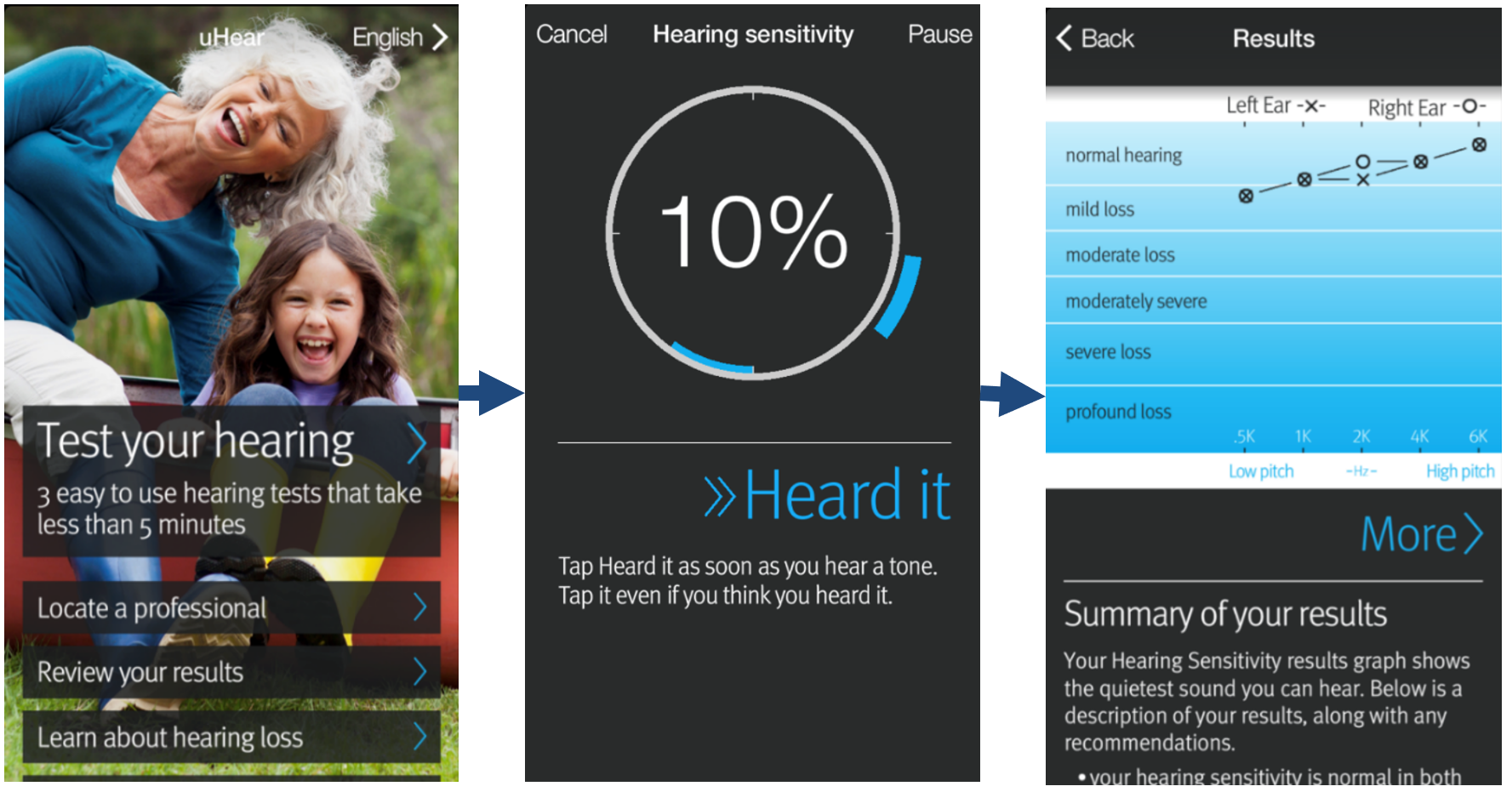}
    \caption{Procedure of pure tone audiometric test on uHear app}
    \label{fig:uHear_screenshot}
\end{figure}

Conventional pure tone audiometry (PTA) was conducted using a ``gold standard'' audiometer (AD629, Interacoustics A/S, Middelfart, Denmark) in a quiet room within the permissible ambient noise limits \cite{ISOAudiometry2010}. Conventional PTA is a manual determination of threshold levels as defined by the American Speech-Language-Hearing Association \cite{ASHA2005}, and is also referred to as the ascending method in ISO 8253-1, or the modified Hughson-Westlake procedure as described in the British Society of Audiology guidelines \cite{BSA2018,ISOAudiometry2010}. It is characterized by its ``5 dB up'' and ``10 dB down'' procedure in determining threshold levels. In contrast, the uHear app is an automated implementation of the modified Hughson-Westlake method with the same starting level of 30 dB HL (Hearing Level) as the conventional PTA. All the thresholds determined by the uHear app were through the same set of devices (iPhone 4S and Earpods, Apple Inc., California, USA). Instructions to conduct the automated audiometric assessment on the uHear app was presented via instructions on a graphic user interface, as shown in \Cref{fig:uHear_screenshot}.

\subsection{Validation procedure}\label{sec:metrics}

The audiograms obtained from both the audiometer and uHear app were graded based on the WHO hearing loss grading system \citep{WorldHealthOrganisation2021}. The grading criteria for ``normal hearing'' requires the hearing threshold (HT) in the better ear to be less than \SI{20}{\decibel HL} (lowered from \SI{26}{\decibel HL} in 2021), and that the HT in the worse ear is less than \SI{35}{\decibel HL}. The HT is determined by the arithmetic mean of hearing levels in dB HL across 4 tested frequencies (0.5, 1, 2, and \SI{4}{\kilo\hertz}). Since the uHear app also assesses the \SI{6}{\kilo\hertz} frequency, it is included along with the 4 frequencies in the WHO grading scheme, denoted as ``WHO+". The suitability of the uHear app would be assessed based on a set of metrics (i.e. accuracy, sensitivity, specificity, and precision) in grading normal hearing via both WHO and WHO+ schemes. 

The accuracy, precision, sensitivity and specificity of the uHear app in determining ``normal hearing'' was computed based on the binary confusion matrix with the audiometer as the reference. Positives (``1") represent the grading of ``normal hearing", whereas negatives (``0") indicate otherwise. The overall accuracy is determined by
\begin{equation}
    \text{Accuracy} = (\text{TP} + \text{TN})/K,
\end{equation}
where TP and TN are ``true positives'' and ``true negatives", and $K$ is the total number of positives and negatives. The sensitivity and specificity are determined by
\begin{equation}
    \text{Sensitivity} = \text{TP} /(\text{TP}+\text{FN}),
\end{equation}
and
\begin{equation}
    \text{Specificity} = \text{TN} /(\text{FP}+\text{TN}),
\end{equation}
respectively. Lastly, the precision is given by
\begin{equation}
    \text{Precision} = \text{TP} /(\text{TP}+\text{FP}).
    \label{eqn:precision}
\end{equation}

\subsection{Data analysis}

Due to the interval nature of the audiogram measurements, non-parametric representations of the Bland-Altman plots were presented to investigate the agreement between both methods \citep{MartinBland1986,Bland1999}. Distribution-free tests and confidence intervals (CI) were computed for the median to examine the bias, and also for the 2.5\textsuperscript{th} and 97.5\textsuperscript{th} percentiles to discover outliers \citep{ashtFay2022}. The accuracy, sensitivity, specificity, and precision of the uHear app in grading normal hearing ability was evaluated in comparison to the audiometer. To determine if there were differences between both methods in grading normal hearing, a McNemar's test was also conducted \citep{Kuhn2019caret}. Finally, the non-parametric Wilcoxon signed rank tests were employed to determine the differences between the methods for each test frequency \citep{RCoreTeam2021}. All data analyses were conducted with the R programming language \citep{RCoreTeam2021} on a 64-bit ARM environment. The data that support the findings of this study are openly available in NTU research data repository DR-NTU (Data) at \href{https://doi.org/10.21979/N9/JQDI6F}{doi:10.21979/N9/JQDI6F}, and the replication code is available on GitHub at \href{https://github.com/ntudsp/douHear}{github.com/ntudsp/douHear}.

\section{Results and Discussion}

% Hopefully the uHear app is not too bad

\subsection{Agreement between the uHear app and the audiometer}\label{sec:agreement}

\begin{figure}[tb]
    \centering
    \includegraphics[width=0.9\textwidth]{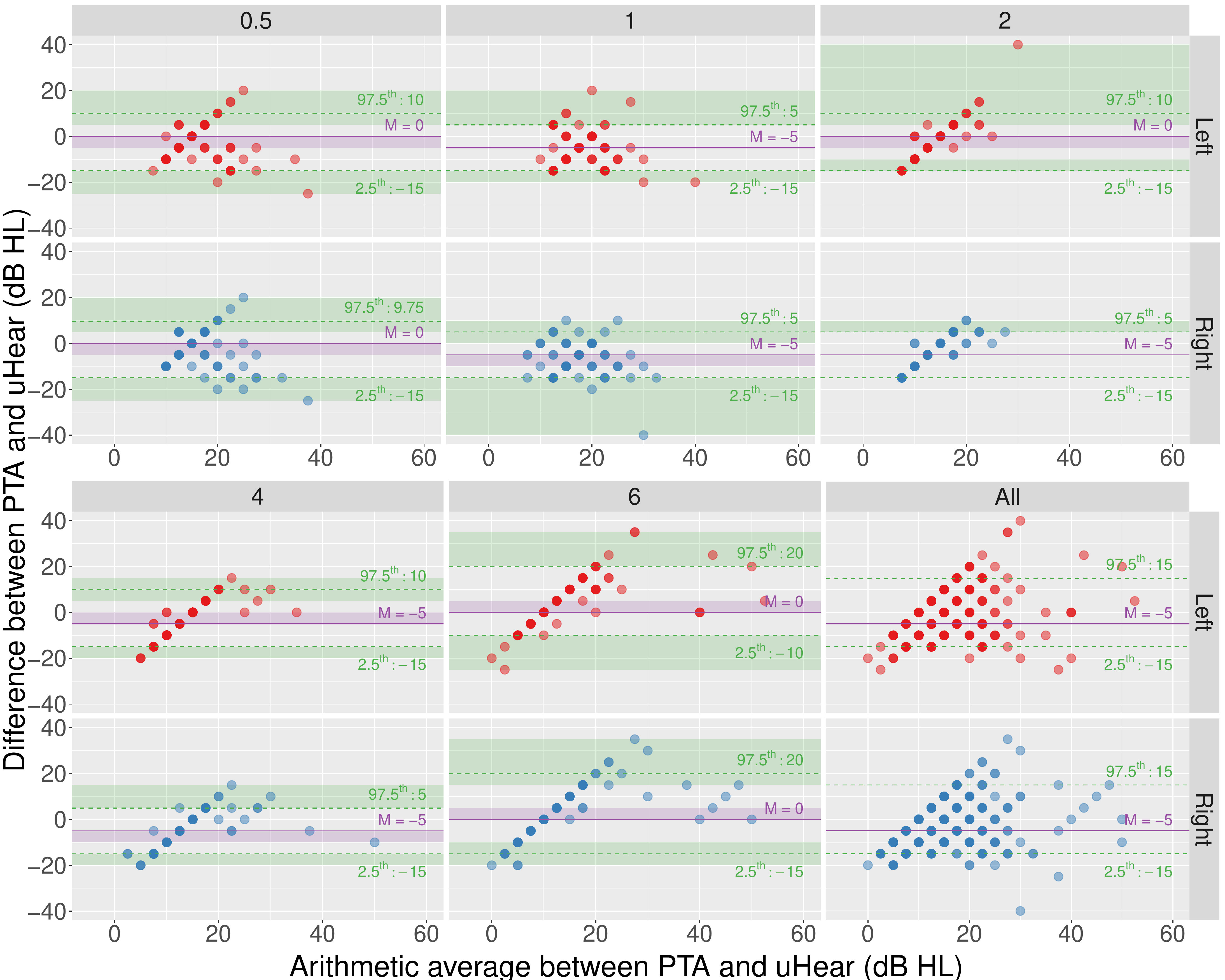}
    % \captionsetup{width=\textwidth}
    \caption{Bland-Altman plot of agreement between the pure tone audiometry and uHear mobile app for measuring hearing level (dB HL) across 0.5, 1, 2, 4, 6 kHz frequencies for both ears. The average bias is estimated by the median of the differences and limits of agreement by the 2.5\textsuperscript{th} and 97.5\textsuperscript{th} percentiles. Distribution-free confidence intervals are based on \SI{95}{\%} confidence level for the upper and lower percentiles (\textcolor{perclr}{$\blacksquare$}) and the median (\textcolor{medclr}{$\blacksquare$}).}
    \label{fig:baplot}
\end{figure}

In the non-parametric implementation of the Bland-Altman plot, the median of the paired differences between audiometer and uHear measurements is used to judge the measurement bias. The lower and upper limits of agreement (LoA) are represented by the 2.5\textsuperscript{th} and 97.5\textsuperscript{th} percentiles, respectively. Since no differences were lower or higher than the \SI{95}{\%} confidence intervals of the lower or upper tails of the respective lower and upper LoAs, no outliers were detected at each frequency across both ears, as shown in \Cref{fig:baplot}. Moreover, the measurements appear to be symmetrically distributed about the medians.

A slight overall bias of about \SI{-5}{\decibel} [(\num{-5}, \num{-5}), \SI{95}{\%} CI] in the uHear app is reflected in the median across both ears when considering differences from all frequencies. The overall bias could be attributed to the \SI{-5}{\decibel} bias at 1 kHz [(\num{-5}, \num{-5}), \SI{95}{\%} CI] in the left ear; and \SI{-5}{\decibel} bias at 1 [(\num{-10}, \num{-5}), \SI{95}{\%} CI], 2 [(\num{-5}, \num{-5}), \SI{95}{\%} CI] and 4 kHz [(\num{-10}, \num{-5}), \SI{95}{\%} CI] frequencies in the right ear, since the equality lines were beyond the confidence intervals in those cases. Hence, there is a greater tendency to obtain a better hearing threshold using the uHear app. This leniency could lead to an increase in false positives.  

\subsection{Normal hearing grading performance of uHear} 

\begin{figure}[tb]
    \centering
    \includegraphics[width=0.9\textwidth]{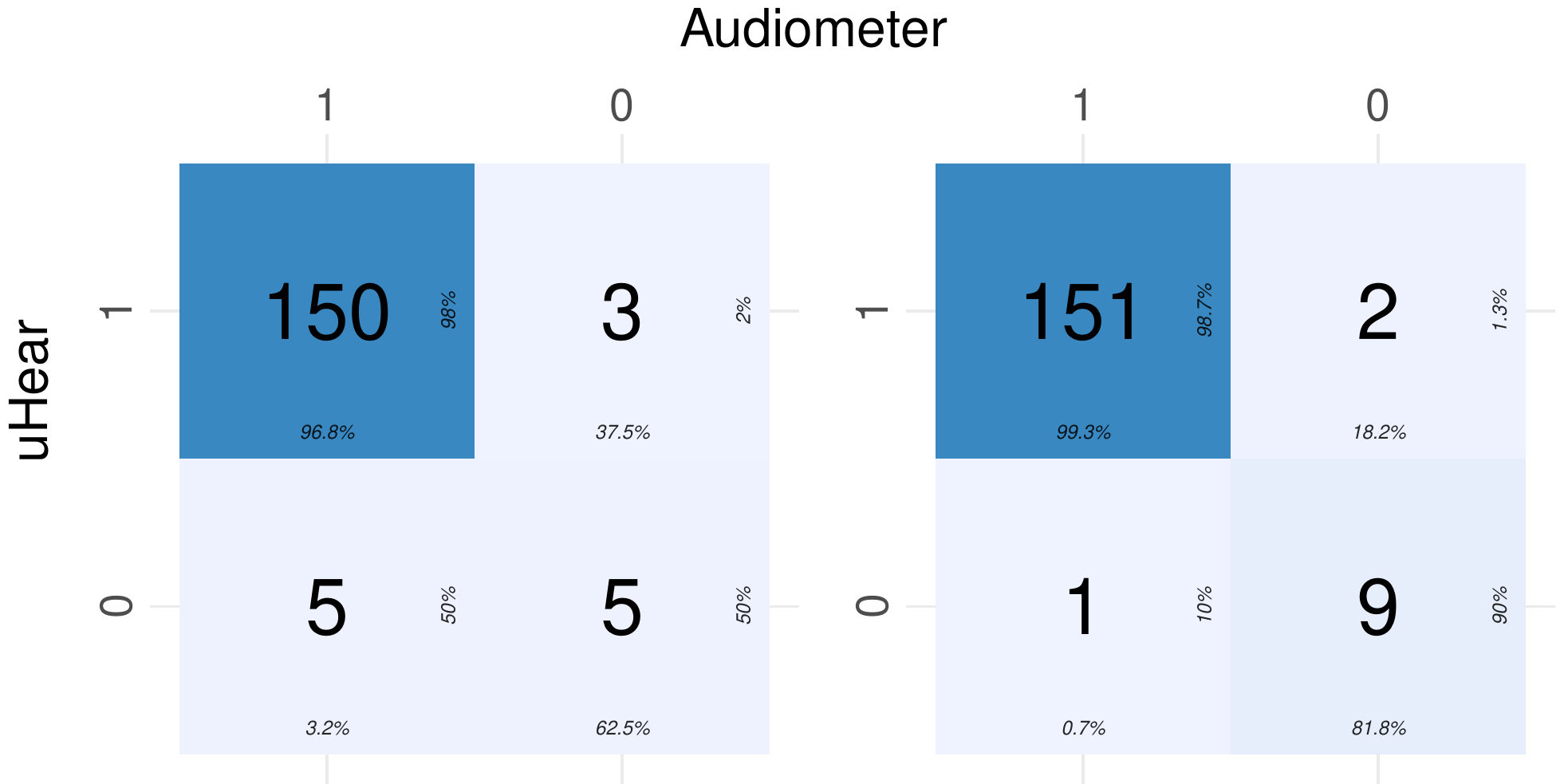}
    % \captionsetup{width=0.9\textwidth}
    \caption{Confusion matrix of the uHear app against the audiometer in grading normal hearing. Normal hearing is indicated as ``1", and ``0'' otherwise, wherein ``11", ``01", ``00", and ``10", respectively indicates ``true positive'' (TP), ``false negative'' (FN), ``true negative'' (TN), and ``false positive'' (FP). Grading is based on the hearing threshold (HT) calculations from WHO (left) and WHO+ (right) schemes, where the latter uses HTs based on the arithmetic mean across all frequencies measured in the uHear app.}
    \label{fig:cfm}
\end{figure}

The normal hearing grading outcomes of all 163 participants across both WHO and WHO+ schemes are presented as confusion matrices, as shown in \Cref{fig:cfm}. At a glance, the number of TNs and FPs were about the same across both schemes, but the number of FPs were substantially lower in the WHO+ scheme. The computed performance metrics outlined in \Cref{sec:metrics} for the uHear app using both WHO and WHO+ grading schemes are summarized in \Cref{tab:cfmetrics}. Across the metrics, the WHO+ scheme appears to be outperforming the WHO scheme, especially in terms of specificity. The low specificity across both schemes provides further evidence of the inability of the uHear app in correctly grading the presence of hearing loss (i.e. TN). Using the McNemar's test (MNT), however, the differences in sensitivity and specificity between the uHear app and the audiometer was not statistically significant in either schemes ($p>0.05$), as shown in \Cref{tab:cfmetrics}.

Different from audiometric diagnostics that detect instances and assess the severity of hearing loss, the aim of the audiometric test for listening experiments is to screen for normal hearing as an inclusion criteria. Whereas FPs would compromise the quality of listening experiments that require participants with normal hearing, it is not detrimental to the listening experiment when an FN occurs (apart from losing potential participants). Hence, for the purpose of screening for normal hearing, precision would be the most important metric to determine the suitability of the uHear app as a low-cost replacement for audiometers, as seen from \Cref{eqn:precision}. Although both WHO and WHO+ grading schemes are comparable in terms of precision, it is worth noting that there was also an overall change in the reference audiometer grading in the WHO (normal: 155 (\SI{95.1}{\%}); not normal: 8) and WHO+ (normal: 152; not normal: 11) schemes. 

%For the ARAUS dataset and SATP that we have conducted, we would rather the uHear app fail participant who actually have no hearing loss (=FN) than the uHear app pass participant who actually have hearing loss (=FP). Therefore, by having a FN just means we will need to get another participant to do the hearing test. But having a FP will compromise the data quality. In other words, the cost of FN will be lower than the cost of FP. Ideally, of all participant that the uHear app allows to pass (=FP+TP), we want as little of them to be FP as possible. Conversely, we want as many of them to be TP as possible, so the correct metric for this measurement of ``goodness'' of uHear app is actually precision (1-(FP)/(TP+FP)). As can be seen, the precision is very high.

%However, note that the limitation of this study is that the results are conditioned on people self-reporting no hearing loss, since we required participants to report on the Google Form for signups that they had no hearing loss before we recruited them for the ARAUS dataset and SATP. Therefore, it is not exactly a super unbiased sample of the population (that would be needed if this were a clinical validation), but it is a reasonable sample of ``the population suitable for soundscape studies". 

\begin{table}[t]
\caption{Accuracy, sensitivity, specificity, and precision, and McNemar's Test \textit{p}-value of the WHO and WHO+ grading criteria for normal hearing. WHO+ includes the measurements at 6 kHz in the grading, which is on top of the 4 frequencies in the WHO grading (i.e. 0.5, 1, 2, and \SI{4}{\kilo\hertz}).}
\centering
\begin{tabularx}{\textwidth}{lc*{3}{>{\centering\arraybackslash}X}c}
\toprule
Grading                     &   
Accuracy (\%; \SI{95}{\%} CI)  &   
Sensitivity (\%)            &   
Specificity (\%)            &   
Precision (\%)              &
MNT \textit{p}-value            \\

\midrule
WHO                     &   
95.09 (90.56, 97.86)        &   
96.77                       &   
62.50                       &   
98.04                       &
0.723                       \\

WHO+                           &   
98.16 (94.72, 99.62)        &   
99.34                       &   
81.82                       &   
98.69                       &
1.000                        \\
\bottomrule
\end{tabularx}
\label{tab:cfmetrics}
\end{table}

\subsection{Audiometric differences between uHear and the audiometer}

To investigate the differences between the audiometer and uHear audiograms at each frequency, the non-parametric Wilcoxon signed rank test (WSRT) was conducted. The choice of the non-parametric WSRT was motivated by the interval nature of the measurements, as well as the symmetry in the measurements as shown in \Cref{fig:baplot}. The difference in audiograms between the audiometer and the uHear app was found to be statistically significant at all frequencies across both ears at \SI{5}{\%} significance level, except at \SI{6}{\kilo\hertz} for the right ear, as shown in \Cref{tab:wrs_results}. 

\begin{table}[t]
\centering
\caption{Wilcoxon signed rank tests, and the pseudomedian and its confidence intervals (\SI{}{\decibel} HL; 95\% confidence level) of the differences between the audiometer and uHear audiograms are computed at each frequency (0.5, 1, 2, 4 and \SI{6}{\kilo\hertz}) across both ears. The null hypothesis is rejected at \textit{p}-value $<0.05$.}
\setlength{\tabcolsep}{12pt}
\begin{tabularx}{\textwidth}{%
    X
    S[table-format=1.1]
    S[table-format=-1.1]
    r@{\hskip0pt}S[table-format=-1.1,table-space-text-pre={[}]
    @{,\hskip-35pt}S[table-format=-1.1]<{{)}}
    S[table-format=1.2e2]}
\toprule
Ear & {Frequency (\si{\kilo\hertz})} & {Pseudomedian} & \multicolumn{3}{c}{Confidence Interval (\SI{95}{\%})\qquad}  & \multicolumn{1}{c}{\textit{p}-value}\\
\midrule
\multirow{5}{*}{\centering\arraybackslash Left}
& 0.5 & -5.0 &\qquad(&-5 & -5& 6.11e-7\\

& 1 & -7.5  &\qquad(&-7.5 & -7.5& 3.37e-19\\

& 2 & -5.0  &\qquad(&-7.5 & -2.5& 7.14e-7\\

& 4 & -5.0  &\qquad(&-7.5 & -5& 1.65e-10\\

& 6 & 2.5  &\qquad(&0 & 5& 1.13e-3\\
    
\midrule
\multirow{5}{*}{\centering\arraybackslash Right}
& 0.5 & -5.0  &\qquad(&-5 & -2.5& 2.14e-6\\

& 1 & -7.5  &\qquad(&-7.5 & -7.5& 4.31e-23\\

& 2 & -5.0  &\qquad(&-7.5 & -5& 1.61e-13\\

& 4 & -7.5  &\qquad(&-7.5 & -7.5& 1.67e-16\\

& 6 & 2.5  &\qquad(&0 & 2.5& 6.82e-2\\
\bottomrule
\end{tabularx}
\label{tab:wrs_results}
\end{table}

The computed pseudomedian \cite{Hollander2015} of the differences between the audiometer and uHear measurements depicts a clear negative bias in the uHear measurements from 0.5 up to \SI{4}{\kilo\hertz} in both ears. Given the narrow confidence intervals, the pseudomedians were almost equivalent to the computed median of the differences in the Bland-Altman plots in \Cref{sec:agreement}, which confirms the symmetry of the measurements (i.e. high precision). Unsurprisingly, the strongest evidence of difference occurred at the 1 and \SI{4}{\kilo\hertz} tones, as also shown in \Cref{fig:baplot}. The distinct differences also corroborates with previous investigations on the uHear and other mobile application based assessment systems \citep{Barczik2018}.

Interestingly, the pseudomedians at \SI{6}{\kilo\hertz} in both ears were near zero and the WSRT was either weakly significant (Left ear; $0.01 < p=1.13\times10^{-3}<0.05$) or not significant (Right ear; $p=6.82\times10^{-2}>0.05$). Hence, the increase in precision with the addition of the \SI{6}{\kilo\hertz} measurement in the grading for normal hearing could be attributed to the similarity of the uHear app to the audiometer at the \SI{6}{\kilo\hertz} test tone. The uHear app was also found to match audiometers more closely in the mid to high frequencies ($\ge$\SI{2}{\kilo\hertz}) in previous studies \citep{Barczik2018}.

\section{Conclusion}

A mobile-based application (uHear) was assessed for its suitability as a screening tool for normal hearing in participants of listening experiments. Due to the availability of a test frequency (i.e. \SI{6}{\kilo\hertz}) beyond those required by the WHO guidelines to grade normal hearing abilities, two grading schemes, namely the WHO and the modified WHO+, were investigated. Based on the assessment with 163 participants, the uHear app was comparable with the audiometer in precision -- the key metric in screening for normal hearing for listening tests -- using both grading schemes. Despite higher overall accuracy, sensitivity, and precision, the proposed WHO+ grading scheme should only be used after assessing the need to include the \SI{6}{\kilo\hertz} test frequency based on the screening requirements.

Nevertheless, the uHear app still falls short as a low-cost diagnostic tool for assessing the severity of hearing impairment due to significant deviation from the audiometer across frequencies. Further caution should be exercised due to the low specificity (i.e. ability in detecting instances of hearing impairment). The results presented were also obtained via a singular device with a now deprecated audio interface (i.e. \SI{3.5}{\milli\meter}) and should be examined for variations in newer models and modern audio interfaces in the future. 

\section*{Acknowledgements}
This research is supported by the Singapore Ministry of National Development and the National Research Foundation, Prime Minister's Office under the Cities of Tomorrow Research Programme (Award No. COT-V4-2020-1). Any opinions, findings and conclusions or recommendations expressed in this material are those of the authors and do not reflect the view of National Research Foundation, Singapore, and Ministry of National Development, Singapore.

\renewcommand{\bibsep}{0pt}
\renewcommand{\bibnumfmt}[1]{\makebox[0.5cm][l]{#1.}}
\bibliographystyle{vancouver}
\bibliography{references_bhan}
%\nocite{*}

\end{document}